\documentclass[aps,twocolumn,prb,amsmath,amssymb,superscriptaddress,longbibliography]{revtex4-2}
\usepackage{graphicx}
\usepackage[colorlinks=true,urlcolor=blue,citecolor=blue,linkcolor=blue]{hyperref}
\usepackage{amsmath}
\usepackage{amssymb}
\usepackage{amsthm}
\usepackage{amsfonts}
\usepackage{enumerate}
\usepackage{subfigure}
\usepackage{centernot}
\usepackage{tikz}
\usepackage{ulem}

\newcommand{\affiliationFDU}{Department of Physics, Fudan University, Shanghai 200433, China}

\newcommand{\affiliationUSST}{School of Physics, University of Shanghai for Science and Technology, Shanghai 200093, China}

\begin{document}

\title{Skyrmion Phase and Non-Fermi Liquid Behavior in Nonsymmorphic Magnetic Weyl Semimetal}

 \author{Xi Luo}
  \affiliation {\affiliationUSST}
 \author{ Yue Yu}
  \thanks{Correspondence to: yuyue@fudan.edu.cn}
   \affiliation{\affiliationFDU}

 \date{\today}

 \begin{abstract}
We investigate the interplay between complex magnetic orders and topological electronic states in nonsymmorphic magnetic Weyl semimetals of the ReAlX family (Re is a rare earth element and X is Si or Ge). We show that a Skyrmion lattice can fundamentally alter the behavior of Weyl fermions, driving the system into a non-Fermi liquid state and producing large, sign-tunable Hall responses. To this end, we construct a lattice model incorporating conduction Weyl fermions coupled to localized magnetic moments via Kondo interaction. Considering a multi-${\bf Q}$ cycloid magnetic configuration that evolves into a Skyrmion lattice under an in-plane Zeeman field, we analyze its profound impact on the band structure through magnetic Brillouin zone and band-folding. Using the Kubo formula, we calculate the conductivity tensor and examine the transport properties in the clean limit. Our results reveal that the Skyrmion lattice induces significant changes in both longitudinal and Hall conductivities. Remarkably, the temperature-dependent resistivity deviates from standard Fermi-liquid behavior ($\rho_{xx}\sim T^2$), exhibiting a non-Fermi liquid power-law scaling ($\rho_{xx}\sim T^\alpha$  with $\alpha$ between 3 and 5). This work provides a unified theoretical framework connecting multi-${\bf Q}$ magnetic textures, Skyrmion physics, and anomalous transport in topological semimetals, bridging the fields of topological magnetism and topological fermions.
 \end{abstract}
\maketitle

\textit{Introduction}—A recently discovered class of non-collinear magnetic Weyl semimetals, ReAlX (Re is a rare earth element and X is Si or Ge) family of materials, has emerged as an ideal platform for exploring the interplay of topological electronic states and strongly correlated physics, owing to their simultaneous possession of topological electronic bands and rich magnetic orders \cite{Hasan2018,White2020,Broholm2021,Philip2021,Tafti2023,Li2023,Hirschberger2025,Hirschberger2026}.  In these Weyl semimetal candidates, both time-reversal and inversion symmetries are broken, therefore, non-zero Heisenberg,  Dzyaloshinskii-Moriya (DM), and  Kitaev-Ising interactions are allowed by  the Ruderman-Kittel-Kasuya-Yosida (RKKY)
interactions mediated by the Weyl fermion \cite{Xiao2015,Askari2015,Zhou2015,Zhou2017,Liang2025}. Meanwhile, the localized 4f electrons of the rare earth elements introduce complex magnetic interactions, giving rise to a variety of novel magnetic structures such as helical, cycloidal, and even Skyrmion lattice orders \cite{Tafti2023,Hirschberger2025,Hirschberger2026}. Of particular interest is the series of significant anomalous transport phenomena experimentally observed in this family. For instance, a large topological Hall effect (THE) has been discovered in the "A-phase" of SmAlSi \cite{Tafti2023}, nearly perfect multi-${\bf Q}$ spin cycloid textures have been observed in GdAlSi \cite{Hirschberger2025}, and longitudinal resistivity scaling of $\sim T^3$ near $10K$, deviating from conventional Fermi liquid theory ($\sim T^2$) or the simple linear behavior of Weyl fermions ($\sim T$) \cite{Vishwanath2012}, has been reported in PrAlGe and LaAlGe \cite{White2020,Philip2021}. The non-Fermi liquid behavior has also been reported in the  Skyrmion lattice phase of MnSi \cite{Pfleiderer2013}.

The physical origin of these anomalous transport behaviors is widely believed to be intimately linked to the complex magnetic orders, topological electronic states, and the strong interaction between them within these materials. On one hand, non-collinear magnetic structures (such as Skyrmions) formed by localized magnetic moments generate a real-space Berry curvature, which acts as an effective magnetic field on conduction electrons, i.e., the THE \cite{Neubauer2009,Hirschberger2019}. On the other hand, the interaction between conduction electrons (here, the Weyl fermions) and localized moments is bidirectional. Localized moments indirectly influence conduction electrons via the RKKY interaction, while the itinerant nature of Weyl fermions, in turn, modulates this interaction \cite{Xiao2015,Askari2015,Zhou2015,Zhou2017,Liang2025}. This can lead to long-range and anisotropic magnetic exchange, potentially stabilizing complex multi-${\bf Q}$ magnetic orders. Furthermore, direct interactions such as the on-site Kondo coupling further entangle the magnetic order with the electronic band structure, potentially inducing band folding, gap opening, or shifting of Weyl nodes, thereby fundamentally altering the Fermi surface topology and low-energy excitation spectrum of the system \cite{Fritz2015,Yu2025}.

Despite the growing body of experimental observations, {a unified theoretical picture remains lacking for how specific magnetic structures, such as the Skyrmion lattice evolving from multi-${\bf Q}$ cycloid orders, quantitatively affect Weyl fermion transport and lead to non-Fermi-liquid behavior in the ReAlX family. Understanding this interplay between real-space topology (magnetic textures) and momentum-space topology (Weyl fermions) is key not only to unraveling exotic quantum phenomena in these materials but also to designing next-generation spintronic and topological devices, such as low-power magnetic memory and topological Hall sensors}\cite{Beidenkopf2022,BatistaPRB2021,Gupta2021,YouAPL2021,Incorvia2023}.

{Motivated by these experimental puzzles, we establish a microscopic theoretical model that reveals how a Skyrmion lattice, formed from multi-${\bf Q}$ cycloid magnetic orders, fundamentally alters the electronic structure and transport properties of Weyl fermions. Our work provides a unified explanation for the large topological Hall effect and the unusual temperature-dependent resistivity observed in the ReAlX family, and identifies the Skyrmion lattice as a key ingredient for realizing non-Fermi liquid behavior.} We first construct an effective lattice Hamiltonian incorporating Kondo coupling, analyzing the magnetic Brillouin zone folding induced by the Skyrmion background and its reshaping of the band structure. Subsequently, by calculating the conductivity tensor via the Kubo formula in the clean limit, we theoretically reproduce the correlation between the sign change of the topological Hall resistivity and a non-trivial Skyrmion number. Furthermore, we elucidate how the temperature scaling of the  {longitudinal} resistivity evolves into a non-Fermi-liquid behavior between $T^3$ and $T^5$ under the influence of magnetic order, providing a theoretical explanation for relevant experimental observations. This study offers a new perspective for understanding the entanglement between magnetic order and topological transport in magnetic topological materials.







\begin{figure}	\includegraphics[width=0.46\textwidth]{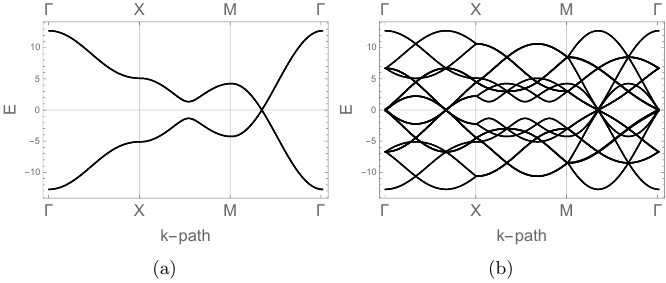}
\caption{(color online) The  band structures of $k_z=0$ plane of (a) $H_0$ and (b) $H$. $m_1=m_2=3$, $u_1=v_2=4$, $v_1=u_2=2$, $v_0=1$, $a=b=0$, and $K=0.2$. In (a), the Weyl nodes are located near $(\pm\frac{2\pi}{3},\pm\frac{2\pi}{3},0)$. In (b), there are extra Weyl nodes emerge near the $\Gamma$ point, $(\pm\frac{2\pi}{3},0,0)$, and $(0,\pm\frac{2\pi}{3},0)$ due to the band folding caused by the multi-${\bf Q}$ magnetic order ${\bf M}$ (\ref{eq3}).}
\label{fig1}	
\end{figure}

\textit{Lattice model}—In the ReAlX-family materials, the nearly commensurate magnetic order couples the Weyl nodes located in the  $k_z=0$ plane \cite{Tafti2023,Hirschberger2025}. It can be described by the following {effective} lattice model \cite{Hasan2018}, 
\begin{eqnarray}
    H_0&=&(m_1+u_1\cos k_x+v_1\cos k_y)\chi_0\sigma_1+t_0\sin k_z \chi_3\sigma_2\nonumber\\
    &+&(m_2+u_2\cos k_x+v_2\cos k_y)\chi_0\sigma_3\nonumber\\
    &+&a(\chi_3\sigma_1-\chi_3\sigma_3)+b\chi_3\sigma_2, \label{eq1}
\end{eqnarray}
    where $\sigma_i$ and $\chi_i$ are Pauli matrices acting on spin and orbital indices, respectively.  To compare with the experiments \cite{Tafti2023,Hirschberger2025}, we choose $m_1=m_2=3$, $u_1=v_2=4$, $v_1=u_2=2$, and $v_0=1$, then there are eight massless Dirac points located at $(\pm\frac{2\pi}{3},\pm\frac{2\pi}{3},0)$ and $(\pm\frac{2\pi}{3},\pm\frac{2\pi}{3},\pi)$ when $a=b=0$ (see Fig. \ref{fig1}a). $a$ and $b$ terms are perturbations that  keep $C_4$ rotation symmetry {at the $k_z=0$ plane}. For $a\neq 0$, a Dirac node is split into two Weyl nodes with opposite chiralities. For $b\neq 0$, inversion and time-reversal symmetries are broken.  {This} moves the Weyl nodes away from $k_z=0,\pm\pi$ planes.  

    Due to the breaking of inversion and time-reversal symmetries in the ReAlX-family materials, there can be non-zero Heisenberg, DM, and Kitaev-Ising interactions between the strong magnetic moments of the $f$ electrons of the rare earth elements and complicated magnetic configurations, such as (anti-)ferromagnet, canted, helical, cycloidal, spiral, and Skyrmion may exist \cite{Tafti2023}. These interactions can be induced through the RKKY interactions mediated by the Weyl fermions \cite{Xiao2015,Askari2015,Zhou2015,Zhou2017,Liang2025}. In turn, these non-trivial magnetic configurations can drastically change the physical properties of the Weyl fermions, such as {the anomalous non-Fermi liquid transport properties observed in ReAlX-family materials \cite{Li2023,Broholm2021,White2020,Philip2021}} while a clear theoretical understanding remains lacking.
    
    To further investigate the interplay between the magnetic configurations and the conducting Weyl fermions, we consider the following effective Hamiltonian,
    \begin{eqnarray}
        H&=&H_0+H_K, \quad H_K=\sum_i K {\bf s}_i\cdot {\bf M}_i, \label{eq2}
    \end{eqnarray}
    where $i$ is the lattice site, $K$ is the on-site Kondo coupling between the conduction Weyl fermion and the magnetic moment of the rare earth elements \cite{Fritz2015,Yu2025}. ${\bf s}_i$ is the spin of the Weyl fermion and ${\bf M}_i$ is the magnetic configuration. In the following, we use an adiabatic approximation when solving the conduction electron problem, i.e., the conduction Weyl fermions are assumed to be fast-moving with respect to the fluctuations of magnetic configuration ${\bf M}$ and ${\bf M}$ is treated as a background. 

    In principle, many complicated magnetic configurations can emerge due to the competitions among Heisenberg, DM, and Kitaev-Ising interactions, let alone the influence of the conducting Weyl fermions. Here, we focus on the Skyrmion configuration and the correlation physics between the Skyrmion lattice and the Weyl fermions. More detailed (finite-temperature) phase diagrams with different types of magnetic orders can be obtained by comparing the Gibbs free energies and will be {presented} in future works.

\textit{Multi-${\bf Q}$ magnetic order, Skyrmion lattice and band structures}—The Skyrmion phase (A-phase) and  {THE} have been observed in helimegnets such as MnSi and Gd$_3$Ru$_4$Al$_1$ \cite{Neubauer2009,Hirschberger2019}. In ReAlX-family, a similar A-phase and THE are reported in SmAlSi \cite{Tafti2023}. Furthermore, perfectly harmonic spin cycloid and multi-${\bf Q}$ textures with nearly  nesting  ${\bf Q}_1=2\pi(\frac{2}{3},-\frac{2}{3},0)$ and ${\bf Q}_2=2\pi(\frac{2}{3},\frac{2}{3},0)$ between the Weyl nodes   are reported  in GdAlSi  \cite{Hirschberger2025}. Since the Skyrmion lattice can be obtained from a linear composition of the helical/cycloid magnetic configurations with  multi-${\bf Q}$ structures \cite{Vishwanath2008}, it is reasonable to assume the emergence of A-phase in ReAlX-family is related to the  multi-${\bf Q}$ behavior of the magnetic configurations. Furthermore, the A-phase in SmAlSi emerges within a regime with finite temperature and in plane external magnetic field \cite{Tafti2023}, which suggests that the Skyrmion lattice is composed by  multi-${\bf Q}$  cycloid ({or one cycloid combined with one helical}) configurations instead of {only} helical ones. The reason is that if the Skyrmion lattice is {only} composed  by multi-${\bf Q}$  helical configurations, then the A-phase should emerge with a finite external magnetic field perpendicular to the ${\bf Q}_1$-${\bf Q}_2$ plane, which is in conflict  with the experimental observations \cite{Tafti2023}. 

Inspired by the experimental observations \cite{Tafti2023,Hirschberger2025}, we consider the following magnetic configuration
\begin{eqnarray}
    &&{\bf M}=(\sin({\bf r}\cdot{\bf Q}_2)/\sqrt{2}+\sin({\bf r}\cdot{\bf Q}_1), \label{eq3} \\ 
  &&  \sin({\bf r}\cdot {\bf Q}_2)/\sqrt{2}+\cos({\bf r}\cdot {\bf Q}_1),\cos({\bf r}\cdot {\bf Q}_2)),\nonumber 
\end{eqnarray}
which is the combination of two cycloid magnetic configurations {propagating} along ${\bf Q}_1$ and ${\bf Q}_2$. The Skyrmion {number} is defined by \cite{Vishwanath2008}
\begin{equation}
    N_{s}=\int d^2{\bf r} \frac{1}{4\pi}\hat{\bf n}\cdot(\partial_x \hat{\bf n}\times \partial_y \hat{\bf n}),
\end{equation}
where $\hat{\bf n}={\bf M}/|{\bf M}|$. To obtain a  nonzero Skyrmion {number}, an external in plane magnetic field is necessary. Therefore, we add a Zeeman term $H_\lambda$ into the system Hamiltonian as the effect induced by the external magnetic field,
\begin{equation}
    H_\lambda=\sum_{a=x,y,z}\lambda_a\sigma_a,
\end{equation}
where $\lambda_a$ is the Zeeman's strength along the $a$-direction. One can effectively  add a term $\frac{2}{K}({\lambda}_x,\lambda_y,\lambda_z)$ into the magnetic configuration ${\bf M}$ (\ref{eq3}) {by} combining $H_K$ and $H_\lambda$ together.

\begin{figure*}	\includegraphics[width=1\textwidth]{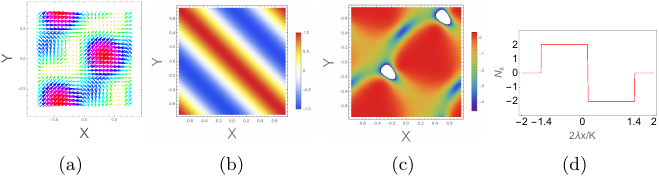}
\caption{{(color online)} The distributions of (a) the $(M_1,M_2)$ and (b) the $M_3$ components of magnetization ${\bf M}$ in a magnetic unit cell. (c) The density distribution of Skyrmion number $N_s$ in a magnetic unit cell.  In (a), (b), and (c), the Kondo coupling $K=0.2$ and the Zeeman strength $\lambda_x=0.1$, $\lambda_y=\lambda_z=0$. (d) The dependence of the Skyrmion number $N_s$ with respect to the in-plane Zeeman field and Kondo coupling in a magnetic unit cell. }
\label{fig2}	
\end{figure*}

We plot a typical Skyrmion lattice in Fig. \ref{fig2}a. By applying {an} in-plane Zeeman field along the $x$-direction, there are two Skyrmions within a magnetic unit cell By choosing $K=0.2$, $\lambda_x=0.1$, and $\lambda_y=\lambda_z=0$ (see Fig. \ref{fig2}a and Fig. \ref{fig2}b), which is consistent with the Skyrmion number $N_s=-2$ within a magnetic unit cell (see Fig. \ref{fig2}d). Without the in-plane Zeeman field, the Skyrmion number is zero when applying Zeeman field in the perpendicular direction alone, which is consistent with the condition for the emergence of A-phase observed in the experiments \cite{Tafti2023}. 

The free Weyl semimetal band structure changes drastically by considering the influence of the Skyrmion lattice through the Kondo coupling $H_K$. {In particular}, the Skyrmion lattice will introduce a magnetic Brillouin zone \cite{Yu2025}. Denote 
\begin{equation}
|\Psi\rangle =\sum_{i,j=1}^3\psi_{i,j}c^\dagger (k_x^0+Qi,k_y^0+Qj,k_z)|0\rangle,     
\end{equation}
where $Q=4\pi/3$ is the norm of the nesting vectors ${\bf Q}_1$ and ${\bf Q}_2$, $ -\frac{\pi}{3}\leq k_x^0,k_y^0\leq \frac{\pi}{3}$, and $-\pi \leq k_z \leq \pi$. Then, the Schr\"odinger equation $H|\Psi\rangle=E|\Psi\rangle$ reduces to the Harper equation \cite{Yu2025,Kohmoto1989,Hatsugai1990,Hatsugai1993} (for more details, see Appendix \ref{app2}).

We plot the band structure in {Fig. \ref{fig1}b} with $K=0.2$. There are extra  Weyl nodes emerging due to the band folding caused by the multi-${\bf Q}$ magnetic order ${\bf M}$. With finite in-plane Zeeman field, the magnetic order ${\bf M}$ can evolve into a Skyrmion lattice (see Fig. \ref{fig2}d), and the Weyl node slightly moves away from its original point for a small Zeeman field. Therefore, the physical properties can change drastically due to the band folding. 





\textit{Conductivity and non-Fermi liquid behavior}—
Experimentally, the transport behaviors of ReAlX-family materials are anomalous and have {attracted} lots of research attention. For instance, large THE is observed in the A-phase of SmAlSi \cite{Tafti2023}, the magnetoresistance {changes} from positive to
negative near 12.8K in CeAlGe \cite{Li2023}, {the electrical resistivity} $\rho_{xx}$ changes drastically in NdAlSi when the background magnetic configuration changes from helical magnet to ferromagnet \cite{Broholm2021}, and $\rho_{xx}\sim T^3$ in PrAlGe and LaAlGe near $T\sim10K$,  which deviates from $T$ linear behavior of Weyl fermion, or $T^2$ behavior of Fermi liquids, or $T^5$ behavior of the electron-phonon interaction, and is believed to be caused by $s-d$ interband scattering \cite{White2020,Philip2021,Vishwanath2012}. Many of the {above-mentioned} anomalous phenomena lack a systematic and theoretic understanding. Here we study the conductivity $\sigma_{ab}$ of the lattice model in the clean limit by the Kubo formula, we find {that the temperature dependence of the longitudinal resistance shows a non-Fermi liquid behavior and behaves differently between the lattice with a non-trivial Skyrmion number and a trivial one,} and the coupling to the multi-${\bf Q}$ cycloid magnetic order drives the system  {deviates from a typical Weyl semimetal}. A more detailed study on the effects of disorder and interactions will be present in future works.

\begin{figure}	\includegraphics[width=0.46\textwidth]{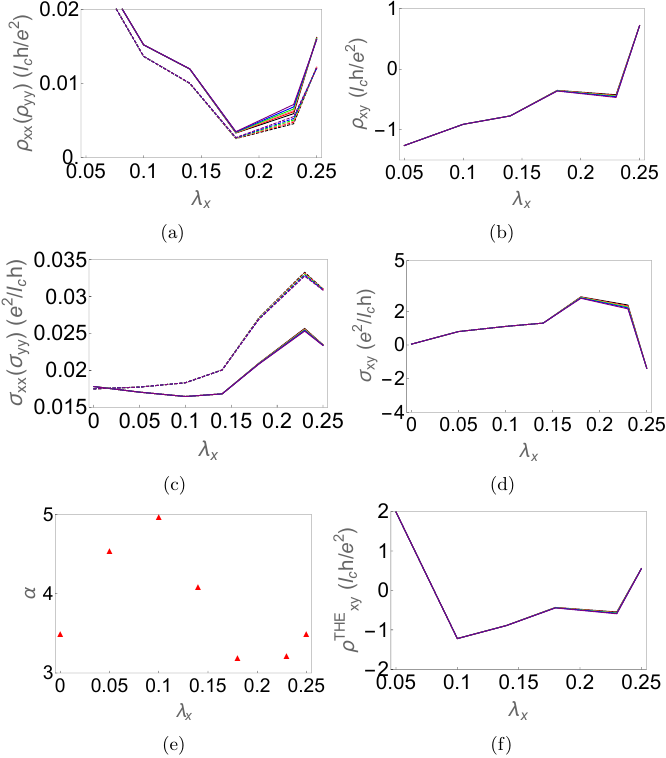}
\caption{(color online)  {(a) and (b) are the resistivity tensor. (c) and (d) are the conductance tensor. (f) is the resistance of THE.} The black, red, pink, green, blue, and purple lines correspond to $T=5K,10K,15K,20K,25K$ and $30K$ respectively. {In (a) and (c), the solid and dashed lines stand for ${xx}$ and ${yy}$ components, respectively.  (e) We plot the fitted $\alpha$ versus $\lambda_x$ with red triangle stands for  the total Hamiltonian $H$ with Kondo interaction (\ref{eq2}).}   In all mini-figures, the external in-plane magnetic field is applied along $x$-direction, {and $l_c$ is the lattice constant in the $z$-direction}. }
\label{fig3}	
\end{figure}

The Kubo formula reads,
\begin{equation}
\sigma_{ab}=\frac{{\rm i}e^2}{h}\int_{BZ} \frac{d^3k}{(2\pi)^3}\sum_{m,n}\frac{f_{nk}-f_{mk}}{E_{nk}-E_{mk}}\frac{\langle nk|v_a| nk \rangle\langle nk|v_b|nk\rangle}{E_{mk}-E_{nk}+i\eta}, \label{kubo}
\end{equation}
where $a,b$ {are the spatial indices}, $E_{nk}$ is the n$th$ band energy of the state $|nk\rangle$ with momentum ${\bf k}$, $f_{nk}$ is the Fermi-Dirac distribution of $E_{nk}$, $v_a=\frac{1}{\hbar}\frac{\partial H}{\partial k_a}$, and $\eta\rightarrow 0^+$. {Furthermore, the conductivity tensor $\sigma_{ab}$ and the electric resistance 
 tensor $\rho_{ab}$ is related by
\begin{eqnarray}
   \rho_{ab}=
   \begin{bmatrix}
       \sigma_{xx} & \sigma_{xy}\\
       -\sigma_{xy} & \sigma_{yy}
   \end{bmatrix}^{-1},
\end{eqnarray}}
We plot the  {conductivity tensor} and the corresponding  {DC resistivity} in Fig. \ref{fig3}. We choose $\eta=0.0001$ and use the classical Monte-Carlo method to numerically evaluate the Kubo formula with the number of the samples $N=90000$ and the random seed being 12345. The results for the total Hamiltonian with nesting multi-${\bf Q}$ cycloid magnetic configuration is devided by 9 due to the band-folding. In SmAlSi, the hopping constant $t\sim 100meV$, and the energy of Weyl node is about $-20meV$ \cite{Tafti2023}, therefore, to simulate the situation in SmAlSi, we choose $a=0.1$, $b=0.2$ and the Fermi level $E_f=0$ in $H_0$ (\ref{eq1}),  which corresponds to $-20meV$ for the Weyl node energy after setting the hopping constant $t=1$.  {From Fig. \ref{fig3}a, the longitudinal conductance $\rho_{xx}$ is about $0.01 l_c h/e^2$ where $l_c$ is the lattice constant along the $z$-direction. In ReAlX, $l_c$ is about 1 $nm$ \cite{Broholm2021,Li2023,Tafti2023}, then $\rho_{xx}\sim 2.5\times10^{-7}\Omega\cdot m$, which is in consistent with the experimental data of SmAlSi \cite{Matusiak2025}. In the Skyrmion phase, the conductance $\sigma_{xx}$ first decreases then increases when the longitudingnal magnetic field increases in  {Fig. \ref{fig3}c}, which is also observed in SmAlSi  \cite{Matusiak2025}. The Hall resistivity changes signs as the magnetic field grows, which indicates a new route to control the Hall currents. Furthermore, due to the anisotropy in the $x$-$y$ plane of the Skyrmion configuration, the $xx$ and $yy$ components in both conductivity and resistivity are slightly different. More details on the resistivity are presented in Appendix \ref{app3}.}


We fit the resistance $\rho_{xx}$ with temperature by the relation $\rho_{xx}=c_1+c_2T^\alpha$, where $c_1$ and $c_2$ are constants and the fitted $\alpha$ are shown in  {Fig. \ref{fig3}e}. Without external magnetic field, $\alpha\sim 3.5$ after considering the nesting multi-${\bf Q}$ cycloid magnetic configuration, and this case is consistent with the experimental observations in PrAlGe and LaAlGe \cite{White2020,Philip2021}. After applying the in-plane magnetic field, $\alpha$ lies between $3-5$, which are all deviates from the Fermi liquid behavior $\rho_{xx}\sim T^2$. {Furthermore, $\lambda_x\sim 0.141$ is the phase boundary between the nontrivial Skyrmion number and the trivial one (see Fig. \ref{fig2}d), we notice that $\alpha>4$ in the Skyrmion phase and $\alpha<4$ in the trivial one which indicates a phase transition.} 

In {Fig. \ref{fig3}f}, we plot the resistance of THE $\rho^{THE}_{xy}$ by subtracting the  contribution of anomalous Hall effect $\rho^A_{xy}$ of the free Hamiltonian $H_0$ (\ref{eq1}) from the total Hall resistance $\rho_{xy}$. We find that the large $\rho^{THE}_{xy}$ fluctuates between  {positive} and negative which indicates the influence of Skyrmion lattice on the transport of the Weyl fermions. {Instead of the THE caused by topologically trivial non-relativistic fermions which is proportional to the Skyrmion number \cite{Nagaosa2004}, the non-linear behavior of THE in {Fig. \ref{fig3}f} implies the competition between the topology of Weyl fermion in momentum space and the topology of Skyrmion in real space. To further investigate the interplay between different topologies,} a more detailed study on the motion of semiclassical wave package of Weyl fermion under the effect of the Berry phase of Skyrmion lattice will be present in future works \cite{niu2010}.

\begin{figure}	\includegraphics[width=0.46\textwidth]{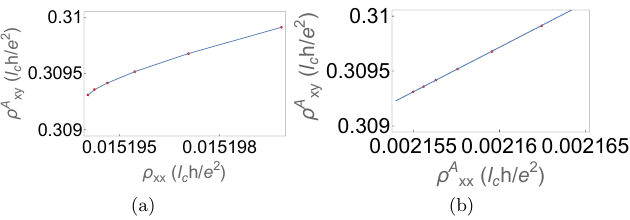}
\caption{(color online) The
scaling behavior of anomalous Hall resistance versus the electric resistance. (a) $\rho_{xy}^A$ vs. $\rho_{xx}$, and (b) $\rho_{xy}^A$ vs. $\rho_{xx}^A$. We choose $\lambda_x=0.1$ in (a) and (b), and the data is obtained from different temperatures. $\rho_{xy}^A$ and $\rho_{xx}^A$ is obtained from the free Hamiltonian $H_0$ (\ref{eq1}) while $\rho_{xx}$ is from the total Hamiltonian $H$ (\ref{eq2}).  }
\label{fig4}	
\end{figure}

{Furthermore, we draw the
scaling behavior of anomalous Hall resistance $\rho_{xy}^A$ versus the longitudinal resistance $\rho_{xx}$ in Fig. \ref{fig4}.} {The free longitudinal conductance $\rho_{xx}^A$ and the anomalous Hall resistance $\rho_{xy}^A$ are obtained from the free Hamiltonian $H_0$ (\ref{eq1}). We find that the relation between $\rho_{xy}^A$ and $\rho_{xx}$ becomes nonlinear (see Fig. \ref{fig4}a), which indicates the nesting multi-${\bf Q}$ cycloid magnetic configuration turns the Weyl semimetal deviates from the high conductivity regime ($\rho_{xy}^A \sim \rho_{xx}^A$, see Fig. \ref{fig4}b) of the free system \cite{Ong2010}. This behavior is also consistent with experiments. In the Skyrmion phase of SmAlSi, the magnitude of resistivity is of order $10^{-7}\Omega\cdot m$ \cite{Matusiak2025} while the resistivity of a typical Weyl semimetal is about $10^{-8} \Omega\cdot m$ in the low temperature regime \cite{Yang2018}. }

\textit{Conclusions}—We have theoretically demonstrated that in nonsymmorphic magnetic Weyl semimetals such as those in the ReAlX family, the multi-${\bf Q}$ cycloid magnetic orders, which can form a Skyrmion lattice, significantly {alter} the electronic and transport properties. The Skyrmion lattice, stabilized by an in-plane Zeeman field, introduces magnetic Brillouin zone, leading to the emergence of additional Weyl nodes and modified band-folding structures.  {The sign-changing behavior of the transverse Hall conductivity  corroborates the intimate link between Weyl fermion and Skyrmion topology}. The temperature scaling of resistivity deviates from conventional Fermi-liquid theory, exhibiting a non-Fermi liquid power-law behavior, consistent with experimental observations in related materials. The scaling relation between anomalous Hall resistivity and longitudinal resistivity reveals that the magnetic order drives the system  {away from a typical Weyl semimetal.} These results establish a coherent picture of how Skyrmion textures interact with Weyl fermions, leading to anomalous transport signatures and non-Fermi liquid behavior. Future work should address the roles of disorder, interactions, and the semiclassical dynamics of wave packets in the Berry phase landscape of Skyrmion lattices.  {Furthermore, it would be possible to verify the proposal of explaining the A-phase in SmAlSi as the multi-${\bf Q}$ cycloid magnetic configurations by using the resonant elastic X-ray scattering method \cite{Hirschberger2025,Hirschberger2026}.} Our study advances the understanding of correlated topological phases and provides a foundation for exploring exotic quantum phenomena {and future design of novel spintronic devices in nonsymmorphic} magnetic Weyl materials.

 \acknowledgements

The authors thank Long Liang for useful discussions. This work is supported by the National Natural Science Foundation of China with Grant No. 12174067.


\section*{DATA AVAILABILITY}

The data that support the findings of this article are not publicly available. The data are available from the authors upon reasonable request.

\newpage

\appendix

\begin{widetext}

\section{Detailed construction of the Harper equation} \label{app2}

Define the partial Fourier transformation,
\begin{eqnarray}
    c^\dagger_{x,y}=\int\frac{d^2k}{(2\pi)^2}c^\dagger_{k_x,k_y} e^{ik_xx+ik_yy}.
\end{eqnarray}
Then 
\begin{eqnarray}
\sum_{x,y}\cos(Q(x+y))c^\dagger_{x,y}c_{x,y}&=&\int\frac{d^2k}{(2\pi)^2}\frac{1}{2}(c_{k_x-Q,k_y-Q}^\dagger c_{k_x,k_y}+c_{k_x+Q,k_y+Q}^\dagger c_{k_x,k_y}),\\
    \sum_{x,y}\cos(Q(x-y))c^\dagger_{x,y}c_{x,y}&=&\int\frac{d^2k}{(2\pi)^2}\frac{1}{2}(c_{k_x-Q,ky+Q}^\dagger c_{k_x,k_y}+c_{k_x+Q,k_y-Q}^\dagger c_{k_x,k_y}),\\
\sum_{x,y}\sin(Q(x+y))c^\dagger_{x,y}c_{x,y}&=&\int\frac{d^2k}{(2\pi)^2}\frac{1}{2i}(c_{k_x-Q,k_y-Q}^\dagger c_{k_x,k_y}-c_{k_x+Q,ky+Q}^\dagger c_{kx,ky}),\\
    \sum_{x,y}\sin(Q(x-y))c^\dagger_{x,y}c_{x,y}&=&\int\frac{d^2k}{(2\pi)^2}\frac{1}{2i}(c_{k_x-Q,k_y+Q}^\dagger c_{kx,ky}-c_{k_x+Q,k_y-Q}^\dagger c_{k_x,k_y}).
\end{eqnarray}
Therefore, each momentum ${\bf k}$ is coupled to ${\bf k} \pm {\bf Q}_1$, and ${\bf k} \pm {\bf Q}_2$ due to the multi-${\bf Q}$ magnetic order ${\bf M}$ (\ref{eq3}). We abbreviate $c_{k_x+Q,k_y+Q'}$ as $c_{Q,Q'}$, and for $Q=4\pi/3$, consider the basis $(c_{0,0},c_{0,Q},c_{0,2Q},c_{Q,0},c_{Q,Q},c_{Q,2Q},c_{2Q,0},c_{2Q,Q},c_{2Q,2Q})^T$, then correspondence between the total Hamiltonian (\ref{eq2}) and the Harper equation reads 
\begin{equation}
    \cos(Q(x+y))\sim\frac{1}{2}
    \begin{bmatrix}
        0 & 0 & 0 & 0 & 1 & 0 & 0 & 0 & 1\\
        0 & 0 & 0 & 0 & 0 & 0 & 0 & 0 & 0\\
        0 & 0 & 0 & 0 & 0 & 0 & 0 & 0 & 0\\
        0 & 0 & 0 & 0 & 0 & 0 & 0 & 0 & 0\\
        1 & 0 & 0 & 0 & 0 & 0 & 0 & 0 & 1\\
        0 & 0 & 0 & 0 & 0 & 0 & 0 & 0 & 0\\
        0 & 0 & 0 & 0 & 0 & 0 & 0 & 0 & 0\\
        0 & 0 & 0 & 0 & 0 & 0 & 0 & 0 & 0\\
        1 & 0 & 0 & 0 & 1 & 0 & 0 & 0 & 0
    \end{bmatrix},\quad
    \cos(Q(x-y))\sim\frac{1}{2}\begin{bmatrix}
        0 & 0 & 0 & 0 & 0 & 1 & 0 & 1 & 0\\
        0 & 0 & 0 & 0 & 0 & 0 & 0 & 0 & 0\\
        0 & 0 & 0 & 0 & 0 & 0 & 0 & 0 & 0\\
        0 & 0 & 0 & 0 & 0 & 0 & 0 & 0 & 0\\
        0 & 0 & 0 & 0 & 0 & 0 & 0 & 0 & 0\\
        1 & 0 & 0 & 0 & 0 & 0 & 0 & 1 & 0\\
        0 & 0 & 0 & 0 & 0 & 0 & 0 & 0 & 0\\
        1 & 0 & 0 & 0 & 0 & 1 & 0 & 0 & 0\\
        0 & 0 & 0 & 0 & 0 & 0 & 0 & 0 & 0
    \end{bmatrix},
\end{equation}
\begin{equation}
    \sin(Q(x+y))\sim\frac{1}{2i}
    \begin{bmatrix}
        0 & 0 & 0 & 0 & 1 & 0 & 0 & 0 & -1\\
        0 & 0 & 0 & 0 & 0 & 0 & 0 & 0 & 0\\
        0 & 0 & 0 & 0 & 0 & 0 & 0 & 0 & 0\\
        0 & 0 & 0 & 0 & 0 & 0 & 0 & 0 & 0\\
        -1 & 0 & 0 & 0 & 0 & 0 & 0 & 0 & 1\\
        0 & 0 & 0 & 0 & 0 & 0 & 0 & 0 & 0\\
        0 & 0 & 0 & 0 & 0 & 0 & 0 & 0 & 0\\
        0 & 0 & 0 & 0 & 0 & 0 & 0 & 0 & 0\\
        1 & 0 & 0 & 0 & -1 & 0 & 0 & 0 & 0
    \end{bmatrix},\quad
    \sin(Q(x-y))\sim\frac{1}{2i}\begin{bmatrix}
        0 & 0 & 0 & 0 & 0 & 1 & 0 & -1 & 0\\
        0 & 0 & 0 & 0 & 0 & 0 & 0 & 0 & 0\\
        0 & 0 & 0 & 0 & 0 & 0 & 0 & 0 & 0\\
        0 & 0 & 0 & 0 & 0 & 0 & 0 & 0 & 0\\
        0 & 0 & 0 & 0 & 0 & 0 & 0 & 0 & 0\\
        -1 & 0 & 0 & 0 & 0 & 0 & 0 & 1 & 0\\
        0 & 0 & 0 & 0 & 0 & 0 & 0 & 0 & 0\\
        1 & 0 & 0 & 0 & 0 & -1 & 0 & 0 & 0\\
        0 & 0 & 0 & 0 & 0 & 0 & 0 & 0 & 0
    \end{bmatrix},
\end{equation}
\begin{equation}
    \sin k_x\sim 
    \begin{bmatrix}
        \sin k_x & 0 & 0 & 0 & 0 & 0 & 0 & 0 & 0\\
        0 & \sin k_x & 0 & 0 & 0 & 0 & 0 & 0 & 0\\
        0 & 0 & \sin k_x & 0 & 0 & 0 & 0 & 0 & 0\\
        0 & 0 & 0 & \sin (k_x+Q) & 0 & 0 & 0 & 0 & 0\\
        0 & 0 & 0 & 0 & \sin (k_x+Q) & 0 & 0 & 0 & 0\\
        0 & 0 & 0 & 0 & 0 & \sin (k_x+Q) & 0 & 0 & 0\\
        0 & 0 & 0 & 0 & 0 & 0 & \sin (k_x+2Q) & 0 & 0\\
        0 & 0 & 0 & 0 & 0 & 0 & 0 & \sin (k_x+Q) & 0\\
        0 & 0 & 0 & 0 & 0 & 0 & 0 & 0 & \sin (k_x+2Q)
    \end{bmatrix},
    \end{equation}
    \begin{equation}
    \sin k_y\sim \begin{bmatrix}
        \sin k_y & 0 & 0 & 0 & 0 & 0 & 0 & 0 & 0\\
        0 & \sin (k_y+Q) & 0 & 0 & 0 & 0 & 0 & 0 & 0\\
        0 & 0 & \sin (k_y+2Q) & 0 & 0 & 0 & 0 & 0 & 0\\
        0 & 0 & 0 & \sin k_y & 0 & 0 & 0 & 0 & 0\\
        0 & 0 & 0 & 0 & \sin (k_y+Q) & 0 & 0 & 0 & 0\\
        0 & 0 & 0 & 0 & 0 & \sin (k_x+2Q) & 0 & 0 & 0\\
        0 & 0 & 0 & 0 & 0 & 0 & \sin k_y & 0 & 0\\
        0 & 0 & 0 & 0 & 0 & 0 & 0 & \sin (k_y+Q) & 0\\
        0 & 0 & 0 & 0 & 0 & 0 & 0 & 0 & \sin (k_y+2Q)
    \end{bmatrix}.
\end{equation}
By diagonalizing the Harper equation, we obtain the band structure of the total Hamiltonian.

\end{widetext}

\section{More discussions on  resistivity} \label{app3}

\begin{figure}	\includegraphics[width=0.46\textwidth]{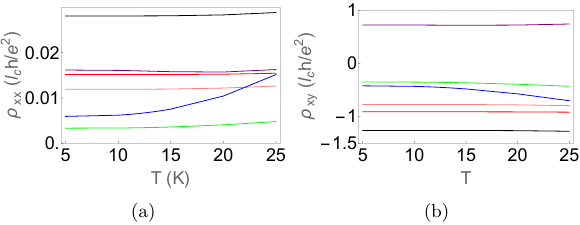}
\caption{(color online)  {The temperature dependence of (a) longitudinal resistivity $\rho_{xx}$ and (b) Hall resistivity $\rho_{xy}$. The black,
red, pink, green, blue, purple lines correspond to $\lambda_x=0.05,0.1,0.14,0.18,0.23,0.25$, respectively. } }
\label{fig5}	
\end{figure}

 {In this appendix, we show the temperature dependence of resistivity tensor with finite $\lambda_x$ in Fig. \ref{fig5}, whose data is in consistent with that in Fig. \ref{fig3} in the main text. We fit $\rho_{xx}= a+bT^\alpha$, and $\alpha$ ranges from 3 to 5 as the red triangle dots in Fig. \ref{fig3}e. }

\begin{figure}	\includegraphics[width=0.46\textwidth]{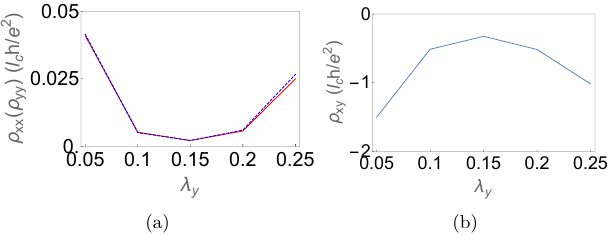}
\caption{(color online)  {(a) 
{longitudinal resistivity $\rho_{xx}$  (red solid line) and $\rho_{yy}$ (blue dashed line),} and (b) Hall resistivity $\rho_{xy}$ at $T=5K$ with $\lambda_y\neq 0$ and $\lambda_x=\lambda_z=0$.} }
\label{fig6}	
\end{figure}

 {Furthermore, we plot the resistivity tensor at $T=5K$ with $K=0.2$, $\lambda_x=\lambda_z=0$ and $\lambda_y\neq 0$ in Fig. \ref{fig6}, namely, the in-plane magnetic field is along the transverse direction. In this case, the Skyrmion number $N_s$  {behaves the same as in Fig. \ref{fig2}d with the phase boundary at $|\lambda_y|\sim 0.141$}. The longitudinal resistance behaves similarly as that in Fig. \ref{fig3}, which decreases first and then increases  as $\lambda_y$ increases. The Hall resistance remains below zero as $\lambda_y$ increases.} 

\end{document}